\documentclass[aps,prl,twocolumn,superscriptaddress,showpacs,floatfix]{revtex4}
\usepackage{graphicx}
\usepackage{latexsym}
\usepackage{stmaryrd}
\usepackage{amsmath}
\usepackage{amssymb}
\usepackage{epstopdf}
\usepackage[colorlinks,linkcolor=red,anchorcolor=red,citecolor=blue]{hyperref}

\begin{document}
\makeatletter
\newcommand{\rmnum}[1]{\romannumeral #1}
\newcommand{\Rmnum}[1]{\expandafter\@slowromancap\romannumeral #1@}
\makeatother

\title{Modified Generalized-Brillouin-Zone Theory with On-site Disorders}

\author{Hongfang Liu }
\affiliation{School of Physical Science and Technology, Soochow University, Suzhou, 215006, China}
\affiliation{Institute for Advanced Study, Soochow University, Suzhou 215006, China}
\author{Ming Lu }\email{luming@baqis.ac.cn}
\affiliation{Beijing Academy of Quantum Information Sciences, Beijing 100193, China}
\author{Zhi-Qiang Zhang }\email{zhangzhiqiangphy@163.com}
\affiliation{School of Physical Science and Technology, Soochow University, Suzhou, 215006, China}
\affiliation{Institute for Advanced Study, Soochow University, Suzhou 215006, China}
\author{Hua Jiang}
\affiliation{School of Physical Science and Technology, Soochow University, Suzhou, 215006, China}
\affiliation{Institute for Advanced Study, Soochow University, Suzhou 215006, China}
\date{\today}

\begin{abstract}
We study the characterization of the non-Hermitian skin effect (NHSE) in non-Hermitian systems
with on-site disorder.
We extend the application of generalized-Brillouin-zone (GBZ) theory to these systems. By proposing a modified GBZ theory, we give a faithfully description of the NHSE. For applications, we obtain a unified $\beta$ for system with long-range hopping, and explain the conventional-GBZ irrelevance of the magnetic suppression of the NHSE in the previous study.
\end{abstract}

\maketitle

\section{Introduction}\label{S1}
Systems described by non-Hermitian Hamiltonians have been attracting intensive attention in recent years \cite{NH1,NH2,NH3,NH4,NH5,NH6,NH7,NH8,NH9,TI1,TI2,TI3,TI4,TI5,TI6,TI7,TI8,TI9,TI10,TI11,TI12,TI13,
TI14,TI15,TI16,Dis1,Dis2,Dis3,Dis4,Dis5,Dis6,Dis7,Dis8,Dis9,Dis10,Dis11,Mag1,Mag2,E,EP1,EP2,EP3,EP4,EP5,
EP6,EP7,EP8,EP9,EP10,EP11,EP12,EP13,PG,QM,NHSE1,NHSE2,NHSE3,NHSE4,NHSE5,NHSE6,NHSE7,NHSE8,NHSE9,NHSE10,
NHSE11,NHSE12,NHSE13,NHSE14,NHSE15,NHSE16,NHSE17,NHSE18,NHSE19,NHSE20,NHSE21,NHSE22,NHSE23,NHSE24,NHSE25,NHSE26,NHSE27}. A large number of interesting phenomena are reported \cite{EP1,EP2,EP3,EP4,EP5,EP6,EP7,EP8,EP9,EP10,EP11,EP12,EP13,PG,QM,NHSE1,NHSE2,NHSE3,NHSE4,NHSE5,NHSE6,
NHSE7,NHSE8,NHSE9,NHSE10,NHSE11,NHSE12,NHSE13,NHSE14,NHSE15,NHSE16,NHSE17,NHSE18,NHSE19,NHSE20,NHSE21,NHSE22,NHSE23,NHSE24,NHSE25,NHSE26}, among which the non-Hermitian skin effect (NHSE) \cite{NHSE1,NHSE2,NHSE3,NHSE4,NHSE5,NHSE6,NHSE7,NHSE8,NHSE9,NHSE10,NHSE11,NHSE12,NHSE13,
NHSE14,NHSE15,NHSE16,NHSE17,NHSE18,NHSE19,NHSE20,NHSE21,NHSE22,NHSE23,NHSE24,NHSE25,NHSE26,NHSE27} has been the focus. The existence of the NHSE indicates that the conventional bulk-boundary correspondence fails \cite{NHSE1,NHSE2,NHSE3,NHSE4,NHSE5}. Meanwhile, the bulk spectrum shows distinct features for the open boundary (OBC) and periodic boundary (PBC) conditions, showing the collapse of the bulk-bulk correspondence (BBC). In order to accomplish the BBC, the generalized-Brillouin-zone (GBZ) theory \cite{NHSE2,NHSE3,NHSE4} introduces a similarity transform for the Hamiltonian, which eliminates the NHSE.

Nevertheless, a recent study \cite{Mag1} showed that the conventional GBZ theory fails to capture the NHSE's features for samples under a magnetic field where the BBC still holds. Loosely speaking, such a model can be considered as the one-dimensional model with an on-site disorder \cite{QusiC}.
Very recently, the modified GBZ theory for the disordered samples was reported \cite{Dis11}, which breaks the limitation of the translational invariance required by the conventional GBZ theory. The essence of the modified GBZ theory is to search the minimum of a polynomial $\mathcal{F}(E,\beta)=|\det[E-\mathcal{H_{PBC}}(\beta)]-\det[E-\mathcal{H_{OBC}}]|$.
However, applying the modified GBZ theory for samples with the on-site disorder is still unreported, which leaves the GBZ irrelevance of magnetic suppression of NHSE unexplained. Thus, there is an urgent need to study the influences of on-site disorder on the GBZ theory.

In this paper, we give the faithful characterization of the NHSE for samples with on-site disorders based on the modified GBZ theory.
We uncover that the transformation coefficient $\beta=\beta_{min}$ determined by the minimum of the polynomial $\mathcal{F}(E,\beta)$ gives an interval instead of a single point.
To unify the description of NHSEs, we demonstrate that the modified GBZ theory also requires the minimization of $|\beta_{min}-1|$.
Based on these considerations, we clarify the applicability of the GBZ theory in several prototypical disordered non-Hermitian models.
In details, a unified transformation coefficient $\beta_{min}$ to achieve the global BBC for disordered samples with long-range hopping is obtained.
A faithful description of NHSEs for samples under the magnetic field is also clarified. Our work removes the ambiguity of the understanding of GBZ theory in the previous studies.

\section{model and method}\label{S2}

We start from the disordered Hatano-Nelson model\cite{HN} with Hamiltonian:
\begin{equation}
\mathcal{H}=\sum_i \varepsilon_i c^{\dagger}_ic_i +t^+c^{\dagger}_i c_{i+1}+t^-c^{\dagger}_{i+1} c_i,
\label{eq1}
\end{equation}
where $c^{\dagger}_i$ ($c_i$) is the creation (annihilation) opperator on the site $i$. $t^{\pm}=(t\pm\gamma)$ represents the nearest-neighbor hopping and $\gamma$ encodes the non-Hermitian strength. We fix $t=1$. $\varepsilon_i \in [-W/2,W/2]$ denotes the on-site disorder \cite{Anderson} with $W$ the disorder strength. The Hamiltonian under OBC and PBC are marked as $\mathcal{H_{OBC}}$ and $\mathcal{H_{PBC}}$, respectively.

Following the modified GBZ theory \cite{Dis11}, we adopt the transformation $t^\pm\rightarrow t^{\pm}\beta^{\pm 1}$. $\beta$ is the transformation coefficient, which gives a quantitative description of NHSE \cite{Dis11,NHSE3,NHSE4}. The transformed Hamiltonian under PBC [$\mathcal{H_{GBZ}}\equiv \mathcal{H_{PBC}}(\beta)$] satisfies \cite{Dis11}:
\begin{align}
\begin{split}
\det[E-\mathcal{H}_{\mathcal{GBZ};N\times N}]=\det[E-\mathcal{H}_{\mathcal{OBC};N\times N}]+f_{\mathcal{PBC}},
\end{split}
\end{align}
where $f_{\mathcal{PBC}}=t^+t^-\det[E-\mathcal{H}_{\mathcal{OBC};N-2\times N-2}]+(t^+)^N\beta^N+(t^-)^{N}\beta^{-N}$. $N$ denotes the size of the Hamiltonian and $E$ is the eigenvalue of $\mathcal{H}_{\mathcal{OBC};N\times N}$. We mark $\mathcal{F}(E,\beta)=|\det[E-\mathcal{H_{PBC}}(\beta)]-\det[E-\mathcal{H_{OBC}}]|$ as follow:
\begin{align}
\begin{split}
\mathcal{F}(E,\beta)&\equiv |f_{\mathcal{PBC}}|=|\mathcal{F}_1+\mathcal{F}_2|,
\label{eqx}
\end{split}
\end{align}
where $\mathcal{F}_1=t^+t^-\det[E-\mathcal{H}_{\mathcal{OBC};N-2\times N-2}]$ and $\mathcal{F}_2=(t^+)^N\beta^N+(t^-)^{N}\beta^{-N}$. For $W=0$, one should notice $\mathcal{F}_1$ is negligible since $\det[E-\mathcal{H}_{\mathcal{OBC};N\times N}]=0$ and $\det[E-\mathcal{H}_{\mathcal{OBC};N-2\times N-2}]\approx0$ \cite{Dis11}. The modified GBZ theory requires the minimum of $\mathcal{F}(E,\beta)=|\mathcal{F}_2|$, which gives rise to $\beta_{min}=\sqrt{|t^-/t^+|}$, consistent with the previous GBZ theory \cite{NHSE3,NHSE4}. Here, $\beta=\beta_{min}$ gives the minimum of the polynomial $\mathcal{F}(E,\beta)$.

\begin{figure}[t]
   \centering
    \includegraphics[width=0.48\textwidth]{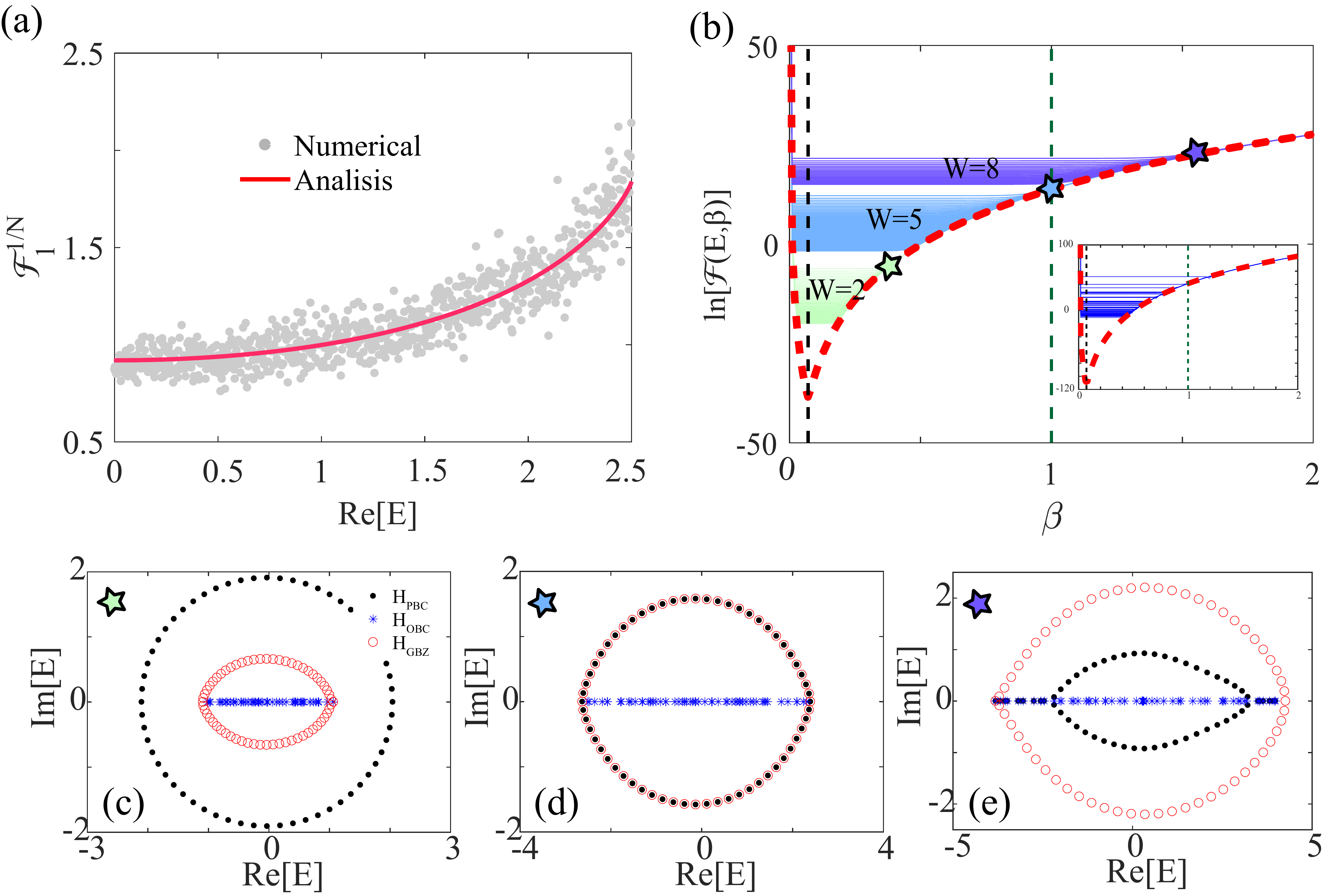}
    \caption{(a) The analytical (red line) and numerical (gray dots) results of $\mathcal{F}^{1/N}_1$ versus the eigenvalue $E$ for $W=5$. (b) $\ln[\mathcal{F}(E,\beta)]$ versus $\beta$ for different eigenvalues with $E\in[-W/2,W/2]$ under disorder strengths: $W=2$ (green), $W=5$ (blue) and $W=8$ (purple). Pentagrams mark the right boundary of the plateaus for $E=\frac{W}{2}$. The red dashed line is the plot of $\ln[|\mathcal{F}_2|]$ for clean samples. Inset is the numerical results for $\mathcal{F}(E,\beta)=|\det[E-\mathcal{H_{PBC}}(\beta)]-\det[E-\mathcal{H_{OBC}}]|$ with $W=8$. (c)-(e) The real part $Re[E]$ versus the imaginary part $Im[E]$ of the eigenvalues $E$ for (c) $W=2$; (d) $W=5$; (e) $W=8$. $\beta$ for $\mathcal{H_{GBZ}}$ is marked in (b). Other parameters are set as $\gamma=0.99$ and $N=60$.}
   \label{f1}
\end{figure}

In the following, we apply the modified GBZ theory for samples with on-site disorders. When disorder is strong enough, we demonstrate that $\mathcal{F}_1$ has considerable influence on achieving the appropriate characterization of the NHSE since $[\mathcal{F}_1(E)]/min[|\mathcal{F}_2(\beta)|]\propto (E-\varepsilon_i)/t$. We have to emphasize that $\mathcal{F}_1$ can be neglected for samples with hopping disorder in our previous study \cite{Dis11}, since the NHSEs are mostly contributed from eigenstates near $E=0$ for considerable disorder. Noticing $[\mathcal{F}_1(E)]/min[|\mathcal{F}_2(\beta)|]\propto E/t\sim0$, it is reasonable to neglect $\mathcal{F}_1$ in our previous study \cite{Dis11}. Besides, for weak disorder, $\mathcal{F}_1$ is always negligible.

For illustration purpose, we first consider $t\sim \gamma$, where the Hamiltonian $\mathcal{H}_{OBC;n\times n}$ is roughly a triangular matrix. The eigenvalue of $\mathcal{H}_{\mathcal{OBC};n\times n}$ can be considered to satisfy $E_{\mathcal{OBC};n}\in\left\{E_i\right\}\approx \left\{\varepsilon_1,\varepsilon_2,\ldots,\varepsilon_i,\ldots,\varepsilon_n\right\}$ with $E_{\mathcal{OBC};n}\in[-W/2,W/2]$.
For $|W/2|>|t^{\pm}|$ and thermaldynamic limit $N\rightarrow\infty$, $\mathcal{F}_1=t^+t^-\det[E-\mathcal{H}_{\mathcal{OBC};N-2\times N-2}]$ can be rewritten as
\begin{align}
\begin{split}
\mathcal{F}_1\approx\prod_{i=1}^{N-2} (E-E_i)t^+t^-\approx\prod_{i=1}^{N} (E-E_i).
\label{eq2}
\end{split}
\end{align}
 $E_i\in E_{\mathcal{OBC};N-2}$ is the eigenvalue of $\mathcal{H}_{\mathcal{OBC};N-2\times N-2}$.
By considering the differences of the eigenvalues between $\mathcal{H}_{\mathcal{OBC};N\times N}$ and $\mathcal{H}_{\mathcal{OBC};N-2\times N-2}$, the analytical formula of $\mathcal{F}_1$ is obtained [see the Appendix for more details]:
\begin{align}
\begin{split}
\mathcal{F}_1=[(E-\frac{W}{2})^{(\frac{1}{2}-\frac{E}{W})}(E+\frac{W}{2})^{(\frac{1}{2}+\frac{E}{W})}e^{-1}]^N.
\label{eq6}
\end{split}
\end{align}
As plotted in Fig. \ref{f1}(a), $\mathcal{F}_1\sim 2^N$ for $E=\frac{W}{2}$, which is comparable with $\mathcal{F}_2$. Thus, $\mathcal{F}_1$ should have distinct influences and cannot be neglected.

In order to better understand the influence of $\mathcal{F}_1$, we plot $\mathcal{F}(E,\beta)$ versus $\beta$ based on our analytical results, shown in Fig. \ref{f1}(b). By neglecting $\mathcal{F}_1$ in Eq.~(\ref{eqx}), the minimum of $\mathcal{F}(E,\beta)=|\mathcal{F}_2|=|(t^+)^N\beta^N+(t^-)^{N}\beta^{-N}|$ and $\beta_{min}$ are $W$ and $E$ independent.
After considering $\mathcal{F}_1$, the global minimum of $\mathcal{F}(E,\beta)=|\mathcal{F}_1+\mathcal{F}_2|$ is still $\beta_{min}=\sqrt{|t^-/t^+|}$. Nevertheless, $\mathcal{F}(E,\beta)$ versus $\beta$ give rise to some plateaus, which is distinct from the result for clean samples [red dashed line in Fig. \ref{f1}(b)]. Moreover, the value of the plateau roughly equals to the global minimum $\mathcal{F}(E,\sqrt{|t^-/t^+|})$. Such a feature is also identified by numerical calculations directly based on Eq. (\ref{eqx}), as shown in the inset of Fig. \ref{f1}(b).
Due to the existence of the plateau for $\mathcal{F}(E,\beta)$, the $\beta_{min}$ should be extended from a single point $\beta_{min}=\sqrt{|t^-/t^+|}$ to an interval $\beta_{min}\in[\Delta^-, \Delta^+]$ for a specific eigenvalue $E$.  Here, $\Delta^\pm$ is determined at the boundary of the plateau.

Since $\beta_{min}\in[\Delta^-, \Delta^+]$ roughly give the same value $\mathcal{F}(E,\beta_{min})$, all the $\beta_{min}$ in the interval can be adopted to realize the BBC with considerable accuracy, and every $\beta_{min}$ gives a correct description of NHSEs.
Nevertheless, to compare the NHSEs for different cases, we demonstrate that the best choices of $\beta_{min}$ should satisfy two key points: (1) {\it it should be in the range $\beta_{min}\in[\Delta^-, \Delta^+]$}, which captures a correct description of NHSEs; (2) {\it $\beta_{min}$ is the one closest to $\beta=1$, i.e., $|\beta_{min}-1|$ has the minimum value.}

Physically, the criterion of BBC ensures \cite{Dis11,NHSE3}:
\begin{equation}
\left\{
\begin{array}{l}
\mathcal{H_{OBC}}\psi_n=E_n\psi_n;\\
\mathcal{H_{PBC}}(\beta_{min})\widetilde{\psi}_n(\beta_{min})=E_n\widetilde{\psi}_n(\beta_{min});\\
\psi_n\approx S\widetilde{\psi}_n(\beta_{min}).
\end{array}\right.
\label{eq7}
\end{equation}
 Here, $S=diag[\beta_{min},\beta_{min}^2,\cdots,\beta_{min}^N]$ is a diagonal similarity transformation matrix. For a specific non-Hermitian sample, the wavefunction $\psi_n$ is determined. In the clean limits, $\widetilde{\psi}_n(\beta_{min})$ is always an extended state.
Conversely, for dirty samples, $\widetilde{\psi}_n(\beta_{min})$ goes from extended to localized by varying $\beta_{min}$ due to on-site disorders. Thus, the corresponding $\beta_{min}$ expands to an interval to restore the original extensibility of $\widetilde{\psi}_n(\beta_{min})$, where $\psi_n\approx S\widetilde{\psi}_n(\beta_{min})$ still holds.
 Notably, the $\beta_{min}$ satisfying $min|\beta_{min}-1|$ gives rise to the very likely extended $\widetilde{\psi}_n(\beta_{min})$, which is close to the clean limits. Besides, when the NHSE is absent, one always has $\beta=1$. Therefore, we suggest adopting the $\beta_{min}$ satisfying $min|\beta_{min}-1|$ as the best depiction of NHSEs.

These characteristics can be identified by $\mathcal{F}(E,\beta)$ and the plot of eigenvalues for different disorder strength $W$, as shown in Figs. \ref{f1}(b)-(e). Based on the proposed theory, one will anticipate the NHSE being destroyed when the right boundary of the plateau [$\Delta^+$ marked by pentagram] crosses the critical value $\beta=1$. Such a feature is confirmed by the spectrums shown in Figs. \ref{f1}(c)-(e).
Generally, the eigenvalues have the NHSE if they form a closed loop with nonzero area in the complex energy plane under PBC (black dots) \cite{NHSE7}.
 When the eigenvalues of $\mathcal{H}_{\mathcal{PBC}}$ and $\mathcal{H}_{\mathcal{OBC}}$ overlap, the corresponding eigenstates are localized due to disorder.

For clarity, we focus on the eigenvalue $E=\frac{W}{2}$.  According to our analytical results shown in Fig. \ref{f1}(b), $W=5$ gives the critical point, and the eigenvalue of $\mathcal{H}_{\mathcal{PBC}}$ and $\mathcal{H}_{\mathcal{OBC}}$ should overlap with $\beta_{min}=1$.
For $W<5$, the BBC requires $\beta_{min}<1$, implying the existence of NHSE for such an eigenvector. It is consistent with the plot in Fig. \ref{f1}(c).
When $W=5$, the eigenvalues of $\mathcal{H_{OBC}}$ and $\mathcal{H_{PBC}}$ [$\beta_{min}=1$] overlap, as shown in Fig. \ref{f1}(d).
By further increasing $W$, the tails of the eigenvalues for $\mathcal{H_{PBC}}$ exist [see Fig. \ref{f1}(e)], and the correlated eigenstates are localized.
However, the BBC for $E=\frac{W}{2}$ is remained available by adopting $\beta_{min}\sim1.5$. In the clean limits, $\beta_{min}\neq1$ predicts the exists of NHSE. It naively gives an inappropriate description of NHSE, where $\psi_n$ and $S\widetilde{\psi}_n(\beta_{min})$ have no NHSEs.
Thus, a faithful characterization of the NHSE should minimize $|\beta_{min}-1|$ for $\beta_{min}\in[\Delta^-, \Delta^+]$. Because, based on the modified BBC theory, the localization features are captured by the plot of $\mathcal{F}(E,\beta)$ at $\beta=1$, as shown in Fig. \ref{f1}(b).

\section{a unified transformation coefficient for disordered samples with next-nearest-neighbor hopping}\label{S3}

In previous section, we find that $\beta_{min}$ is extended from a single point to an interval for achieving the BBC for disordered samples. As an application of such a feature, a unified transformation coefficient $\beta$ is available for samples with long-range hopping even though it does not exist under the clean limits.

\begin{figure}[t]
   \centering
    \includegraphics[width=0.48\textwidth]{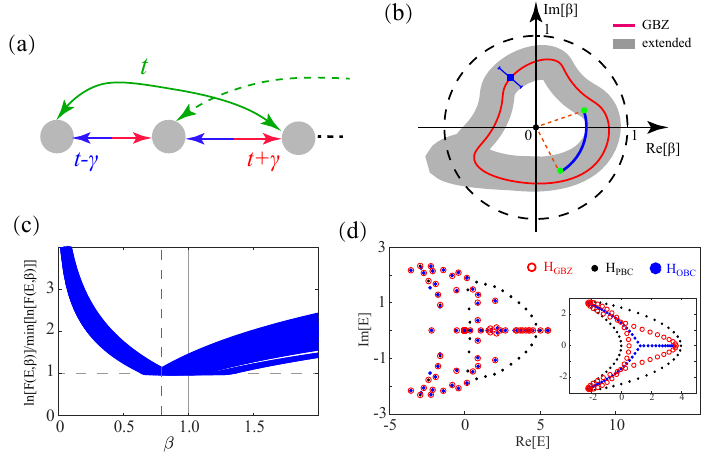}
    \caption{(a) Schematic diagram of the
     model with next-nearest-neighbor hopping (green lines). (b) The schematic plots of $Re[\beta_{min}]$ versus $Im[\beta_{min}]$ for clean (red solid line) and disordered samples (area in gray).
     The disorder effect leads to the broadening of $\beta_{min}$ marked by the error-bar. Thus, a unified $|\beta|$ is available as marked in blue solid line.
      (c) $\ln[\mathcal{F}(E,\beta)]/\min[\ln[\mathcal{F}(E,\beta)]]$ versus $\beta$ for disorder strength $W=7$. The red dashed line presents the overlap of $\beta_{min}$ for different eigenvalues. (d) $Re[E]$ versus $Im[E]$ under PBC (black dots), OBC (blue dots) and GBZ [PBC with $\beta=0.77$ (red dashed line in (c))] (red circles). Inset is the plot for clean samples with $\beta\approx0.854$, which gives rise to the BBC for $E\approx-0.2\pm 1.6i$. We fix $t=1$, $\gamma=1.4$ and $N=60$.}
   \label{f2}
\end{figure}

We focus on samples with next-nearest-neighbor hopping
$$\mathcal{H}=\sum_i \varepsilon_i c^{\dagger}_ic_i +t^+c^{\dagger}_i c_{i+1}+t^-c^{\dagger}_{i+1} c_i+tc^{\dagger}_i c_{i+2}+tc^{\dagger}_{i+2} c_{i},$$
as shown in Fig. \ref{f2}(a).
When disorder is absent ($W=0$), the $\beta_{min}$ can be accurately determined, and the plot of $Re(\beta)$ versus $Im(\beta)$ forms a close loop as shown in the red curve in Fig. \ref{f2}(b). Clearly, a unified transformation coefficient $|\beta_{min}|$ to achieve the global BBC does not exist \cite{NHSE2,NHSE3}, where $\beta_{min}$ is $E$-dependent.
For a typical energy $E=E_0$, $\beta_{min}(E_0)$ only ensures the BBC for such a state [see the inset of Fig. \ref{f2}(d) as an example].
Thus, a global BBC cannot be obtained by simply utilizing a specific $\beta_{min} (E)$ for the clean samples.

Nevertheless, after considering disorder effects, the broadening of $\beta_{min}$ will significantly alter the plot of $Re(\beta)$ versus $Im(\beta)$.
To identify the unified transformation coefficient, we calculate the $\mathcal{F}(E,\beta)$ versus $\beta$ for each eigenvalue $E$. As shown in Fig.\ref{f2}(c), the interval of $\beta_{min}(E)$ is determined by requiring $\ln\{\mathcal{F}(E,\beta_{min})\}/\min\{\ln[\mathcal{F}(E,\beta_{min})]\}\approx1$. Remarkably, all the $\beta_{min}(E)$ intersect at a single point, i.e. $\cap_{i\in E}\beta_{min}(i)=0.77$.
 Since $\beta_{min}(E)$ ensures the BBC for the eigenvalue $E$, the transformed Hamiltonian $\mathcal{H_{GBZ}}$ with $\beta=0.77$ will capture all the eigenvalues under OBC [see Fig. \ref{f2}(d)]. The interplay between disorder effects and the BBC unveils one of the exotic properties of non-Hermitian systems.


\section{ application of the modified GBZ theory under magnetic field}\label{S4}

In this section, we apply the modified GBZ theory to systems with magnetic field.
The Hamiltonian reads:
\begin{align}
\begin{split}
\mathcal{H}=&\sum_{x,y}t^\pm c^{\dagger}_{x,y}c_{x\pm\delta_x,y}+t_ye^{\pm i\phi x}c^{\dagger}_{x,y}c_{x,y\pm\delta_y}.
\end{split}
\end{align}
 $\phi$ represents the magnetic flux. Recently,
Lu, {\it et. al.} and Shao, {\it et. al.} both noticed that the strength of NHSE is significantly suppressed \cite{Mag1,Mag2} by increasing $\phi$.
However, the strength of NHSE under the magnetic field cannot be correctly described by the conventional GBZ theory \cite{Mag1}. The conventional GBZ theory seems to indicate that the NHSE is irrelevant to the magnetic field.
Since such a model is roughly equivalent to a one-dimensional model with the on-site disorder \cite{QusiC}, we next elucidate that a faithful description of NHSE is still available based on the proposed modified GBZ theory.

\begin{figure}[t]
   \centering
    \includegraphics[width=0.48\textwidth]{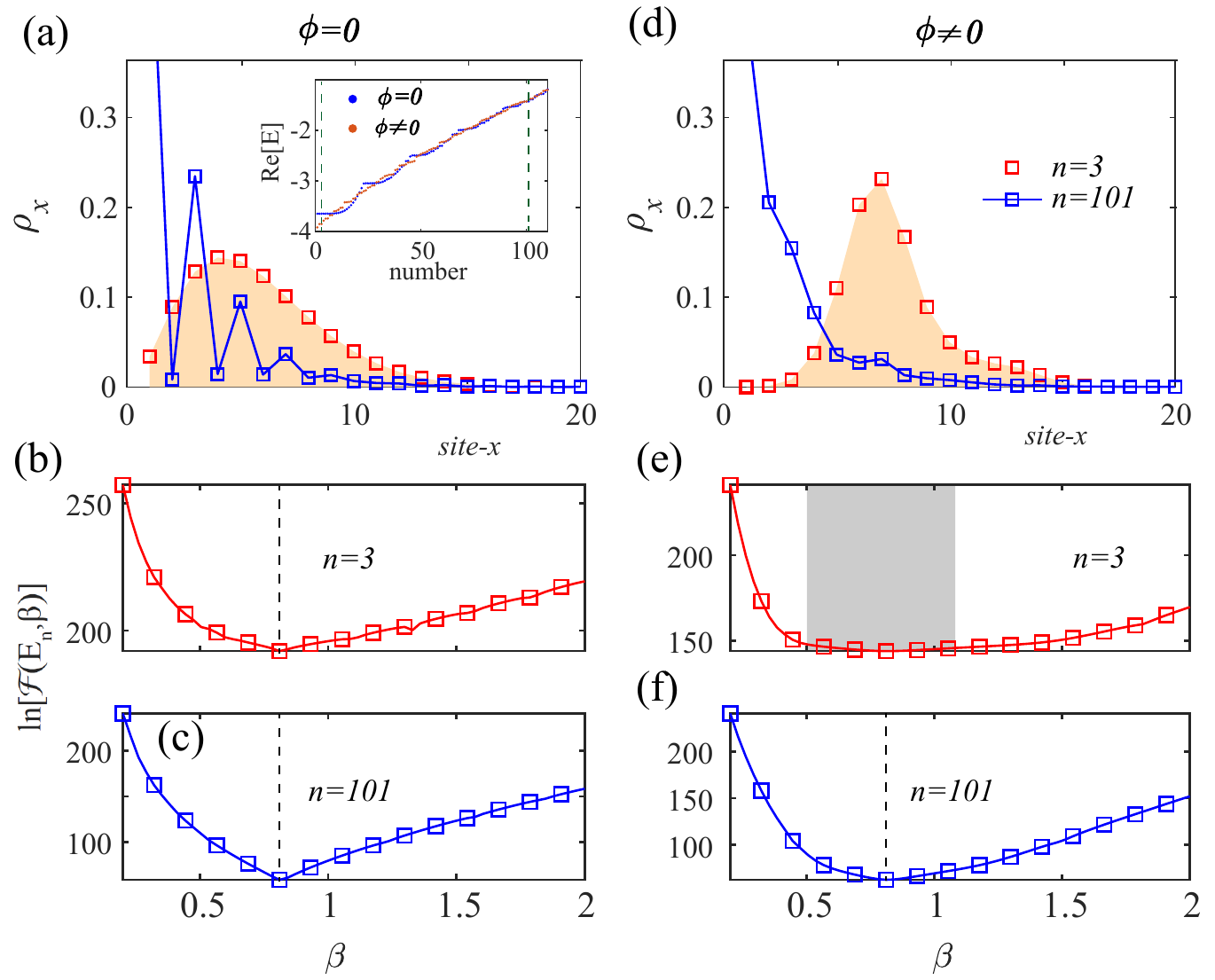}
    \caption{(a) The plot of eigenstates $\rho_{x;n}=\sum_{y}|\psi_n(x,y)|^2$ versus site $x$ for $n=3$ and $n=101$ with $\phi=0$. Inset plots the real eigenvalues ($Re[E]$) for $\phi=0$ (blue dots) and $\phi=2\pi/(N-1)$ (red dots), which determines the choices of $n$. (b) and (c) $\mathcal{F}(E_n,\beta)$ versus $\beta$ for $n=3$ and $n=101$, respectively. (d)-(f) are the same with (a)-(c), except $\phi=2\pi/(N-1)$. We set $t^+=1.2$, $t^-=0.8$, $t_y=1$ and $N_x=N_y=N=20$.}
   \label{f3}
\end{figure}

The eigen-equation gives rise to $\mathcal{H}\psi_n=E_n\psi_n(x,y)$.
$E_n$ stands for the $n_{th}$ eigenvalue shown in the inset of Fig. \ref{f3}(a). For clarity, we take $n=3$ and $n=101$ as two typical examples.
To unveil the universality of the modified GBZ theory under magnetic field, we pay our attention to the evolution of $\mathcal{F}(E_n,\beta)$ versus $\beta$. When magnetic field is absent, the eigenvectors for $n=3$ and $n=101$ should both show the NHSE's features, where a single point of $\beta_{min}\sim 0.81$ is obtained based on $\mathcal{F}(E_n,\beta)$ [see Figs. \ref{f3}(b)-(c)].  After considering the influence of $\phi$, we notice that $\mathcal{F}(E_n,\beta)$ gives a plateau for $n=3$ with $\beta_{min}\in[0.5,1.1]$. Based on the modified GBZ theory, the $\beta_{min}=1$ should be adopted to characterize the NHSE, and the NHSE for $n=3$ should be destroyed. On the contrary, a single point with $\beta_{min}\sim 0.81$ is still available for $n=101$.
These results are consistent with the plot of eigenvectors in Figs. \ref{f3}(a) and (d). The NHSE for $n=3$ is destroyed with $n=101$ unaffected by increasing $\phi$, which manifests the magnetic-field-suppressed NHSE.

We close this section by clarifying why the GBZ theory fails to describe the NHSE. The modified GBZ theory considers the influence of $\mathcal{F}_1$ on the polynomial $\mathcal{F}(E,\beta)$, which smoothes its sharp dip. Nevertheless, such a process leaves the global minimum of $\mathcal{F}(E,\beta)$ unaffected. The applied GBZ theory in the previous study \cite{Mag1} only concentrated on the global minimum, which is almost unaffected by the magnetic field [see Fig. \ref{f3}(e), the global minimum still gives $\beta\sim0.8$]. However, the plateau of $\mathcal{F}(E_{n=3},\beta)$ suggests that both $\beta\in[0.5,1.1]$ can capture the minimum of $\mathcal{F}(E_{n=3},\beta)$ and the BBC with high accuracy. In short, a faithful description of NHSE should pay attention to not only the BBC, but also the additional restrictions of the modified GBZ theory.

\section{summary}\label{S6}

In summary,
we found the minimum of $\mathcal{F}(E,\beta)$ gives an interval instead of a single point, which eases the realization of BBC.
 Due to the extended choices of $\beta_{min}$, the unified transformation coefficient $\beta$ for samples with long-range hopping can be obtained.
  To compare the NHSEs for different cases and eliminate the ambiguous, the strength of NHSEs is unified to the minimum of $|\beta_{min}-1|$. Notably, the modified GBZ theory under strong on-site disorders should concentrate on two key points: (1) {\it the transformation coefficient $\beta=\beta_{min}$ should ensure the correctness of BBC by minimizing $\mathcal{F}(E,\beta)$}; (2) {\it $|\beta_{min}-1|$ should also be minimized}.
 Finally, we clarified the paradox of magnetic-irrelevant NHSEs with the help of the modified GBZ theory.
  Our work deepens the understanding of the characterization of NHSE for samples with on-site disorder and extends the application of the GBZ theory.

\section{ACKNOWLEDGEMENT}
We are grateful for the valuable discussions with Qiang Wei, Shuguang Cheng and Jing Yu.
This work was supported by the National Basic Research Program of China (Grant No. 2019YFA0308403), NSFC under Grant No. 11822407 and No. 12147126, and a Project Funded by the Priority Academic Program Development of Jiangsu Higher Education Institutions.

\section{APPENDIX: The Derivation of Eq. (\ref{eq6})}\label{A}
We give the derivation of the analytical formula of $\mathcal{F}_1$. For Eq. (\ref{eq2}), we mark $t^{+/-}\rightarrow(E-E_i)\neq0$ since there is a high probability $t^{+/-}\sim(E-E_i)$. Here, we suppose $E_i$ and $E$ are the eigenvalue of $\mathcal{H}_{\mathcal{OBC};N-2\times N-2}$ and $\mathcal{H}_{\mathcal{OBC};N\times N}$, respectively. One should also notice that the eigenvalue of $\mathcal{H}_{\mathcal{OBC};N\times N}$ is also approximately the eigenvalue of $\mathcal{H}_{\mathcal{OBC};N-2\times N-2}$, and only a slight deviation $\delta=|E_{\mathcal{OBC};N}-E_{\mathcal{OBC};N-2}|\rightarrow 0$ exists.
By considering $\delta$, the analytical formula of $\mathcal{F}_1$ is obtained
\begin{align}
\begin{split}
&\ln(\mathcal{F}_1)\approx\sum_{i=1}^N\ln(E-E_i)\\
&=\lim_{\delta \rightarrow 0}\frac{N}{W} [\int_{-\frac{W}{2}}^{E-\delta}\ln(E-x)dx+ \int_{E+\delta}^{\frac{W}{2}}\ln(E-x)dx ]\\
&=\ln\{ [(E-\frac{W}{2})^{(\frac{1}{2}-\frac{E}{W})}(E+\frac{W}{2})^{(\frac{1}{2}+\frac{E}{W})}e^{-1}  ]^N\}.
\end{split}
\end{align}
In the above deviation, we require $E_i \rightarrow x$ and $|E-E_i|>\delta$ for arbitrary $E_i\in[-\frac{W}{2},\frac{W}{2}]$.

\end{document}